\documentclass[12pt,preprint]{aastex}
\shorttitle{Ca II Emission from ER~Vul}
\shortauthors{Shkolnik et al.}

\begin{document}

\title{Investigating Ca II emission in the RS CVn binary ER~Vulpeculae using the Broadening Function Formalism}

\author{Evgenya Shkolnik}
\affil{NASA Astrobiology Institute, University of Hawaii at Manoa, 2565 McCarthy Mall, Rm. 213, Honolulu, HI 96822, U.S.A.}
\email{shkolnik@hawaii.edu}

\author{Gordon A.H. Walker\altaffilmark{1}}
\affil{1234 Hewlett Place, Victoria, BC, Canada, V8S 4P7}
\email{gordonwa@uvic.ca}

\author{Slavek M. Rucinski\altaffilmark{1}}
\affil{David Dunlap Observatory, University of
Toronto, P.O.Box 360, Richmond Hill, ON, Canada L4C 4Y6 }
\email{rucinski@astro.utoronto.ca}
%\and

\author{David A. Bohlender and Tim J. Davidge}
\affil{National Research
Council of Canada, Herzberg Institute of Astrophysics\\ 5071 West
Saanich Road Victoria, BC, Canada V9E 2E7}
\email{david.bohlender@nrc.ca, tim.davidge@nrc.ca}

\altaffiltext{1}{Visiting Astronomer, Canada-France-Hawaii Telescope,
operated by the National Research Council of Canada, the Centre
National de la Recherche Scientifique of France, and the University of
Hawaii.}

\begin{abstract}

The synchronously rotating G stars in the detached, short-period (0.7 d), partially eclipsing binary, ER Vul, are the most chromospherically active solar-type stars known. We have monitored activity in the Ca II H \& K reversals for almost an entire orbit.  Rucinski's Broadening Function Formalism allows the photospheric contribution to be objectively subtracted from the highly blended spectra. The power of the BF technique is also demonstrated by the good agreement of radial velocities with those measured by others from less crowded spectral regions. In addition to strong Ca II emission from the primary and secondary, there appears to be a high-velocity  stream flowing onto the secondary where it stimulates a large active region on the surface 30 $-$ 40$^\circ$ in advance of  the sub-binary longitude. A model light curve with a spot centered on the same longitude also gives the best fit to the observed light curve. A flare with $\sim$13\% more power than at other phases was detected in one spectrum. We suggest ER Vul may offer a  magnified view of the more subtle chromospheric effects synchronized to planetary revolution seen in certain `51 Peg'-type systems.

\end{abstract}


\section{Introduction}
RS Canum Venaticorum systems (RS CVn) are (normally) contact binaries with two active late-type stars having orbital periods
between 2 and 20 days (Hall 1992).
Both components have strong chromospheric, transition region and coronal
emission. This is generally assumed to be the result of a powerful dynamo action caused by rapid rotation and deep
convection zones leading to  magnetic activity several orders of magnitude greater than
the Sun's (Vogt et al.~1999).

ER~Vulpeculae is a bright, well-studied RS CVn system which
we chose to test our observation that certain
`51~Peg'-type planets induce detectable chromospheric activity on their parent
stars through magnetic interaction (Shkolnik et al.~2003, 2004). We observed Ca II H (3967 \AA) and  K (3933 \AA) emission as the primary optical indicators of chromospheric activity because the stellar Ca II H and K photospheric absorption lines are so broad and deep that the reversals are seen in high contrast against the
shallow stellar core ($\sim$~10\% of the continuum). ER Vul's short orbital period, brightness (V = 7.4), and high level of chromospheric emission makes it an ideal target for observing phase-dependent variability.

This tidally-synchronized binary is unusual in that it consists of two solar-type
stars orbiting each other with a relatively short period of 0.698 days.  The
stars are nearly identical (G1-2~V and G3~V) with equal radii (1.07
$R_{\odot}$) and a mass ratio of 0.947. They are only 3.97 $R_{\odot}$ apart
in a circular orbit but they are not in contact.  The orbital inclination of 67$^{\circ}$, this configuration
leads to partial eclipses of approximately 12\% (Gunn \& Doyle 1997). The
projected rotational velocities of the primary and secondary components are
($v$sin$i$)$_{P}= 97.4 \pm 1.3$ km~s$^{-1}$ and ($v$sin$i$)$_{S}= 96.7 \pm
2.2$ km~s$^{-1}$ (Duemmler et al.~2003).  This rapid rotation causes very
broad and blended features in the double-lined spectrum.

ER Vul has been studied  in almost every spectral region to discover, map and understand its intense stellar and circumstellar activity. Piskunov (1996) used Doppler
imaging to map the surfaces of both stars from 12 high-resolution spectra centered on 6440 \AA\/ at the
Nordic telescope. He identified hot spots on both components at sub-stellar
longitudes, which he presumed to be due to reflection.

Duemmler et al.~(2003) also report phase-dependent activity on ER~Vul
from asymmetry in the Ca II IRT line at 8662 \AA. They measured
deviations of $\sim$10 km~s$^{-1}$ from the radial velocity (RV) curve at
phases when a star is receding from the observer.  They suggest
that this is caused by localized excess emission near the sub-binary point on
the star. Alternatively, a weakening of the emission at opposite orbital
phases could produce  the same effect. They measured a
higher level of activity on the cooler, secondary component, which was already known to be globally more active than the primary (Gunn \& Doyle 1997;
\c{C}akirli et al.~2003).

VLA radio observations by Rucinski (1992, 1998) revealed a complex, relatively large radio flux variability but having no clear correlation with the orbital period.  This may indicate a structured radio corona. At the other extreme, EXOSAT observations exhibit a homogeneous X-ray corona (White et al.~1987).  The lack of any eclipses in these two wavelengths can be interpreted as either the active regions are at high latitudes or the regions of X-ray and radio emission extend
quite far beyond the binary system, perhaps in the form of huge coronal loops.

The classic chromospheric diagnostics, Ca II H \& K, were looked at by Montes et al.~(1996) and
Fernandez-Figueroa et al.~(1994) as part of a large survey of chromospherically active binaries. However, their
data were too few and sporadic to make any claims of phase-dependent activity. In this paper, we present 42 new Ca II H \& K observations with nearly complete orbital phase coverage. The details of the observations are discussed Section~\ref{spectra}. In Section~\ref{BFs}, we present the broadening functions (BFs) determined from the spectra with the derived RVs and the orbital parameters. The Ca II emission is isolated using the BFs and its origins are discussed in Section~\ref{CaII}. In Section~\ref{photometry}, we model the photometric lightcurve obtained concurrently with the spectroscopic observations--.

\section{The ER~Vul Spectra \label{spectra}}

Ca II H \& K spectra were taken at the Canada-France-Hawaii Telescope (CFHT) with the Gecko \'{e}chellette spectrograph fiber fed by CAFE (CAssegrain
Fiber Environment) from the Cassegrain to Coud\'{e} focus (Baudrand \&
Vitry 2000). Spectra were centered at 3947 \AA\/ which was isolated by a
UV grism (300 lines mm$^{-1}$) with $\simeq$~60~\AA\/ intercepted by the
CCD. The dispersion was 0.0136 \AA\/ pixel$^{-1}$ and the 2.64-pixel FWHM
of the thorium-argon (Th/Ar) lines corresponded to a spectral resolution
of R = 110,000. The detector was a back-illuminated EEV CCD (13.5
$\mu$m$^{2}$ pixels, 200 $\times$ 4500 pixels) with spectral dispersion
along the rows of the device.
The B magnitude of ER~Vul is 7.93 allowing a S/N of $\sim$
180 pixel$^{-1}$ (or 1500 \AA$^{-1}$) to be achieved in an exposure time of 1200 s. We used the same set-up for ER Vul as we did for the `51-Peg' systems presented in Shkolnik et al.~(2003) where the reduction details can be found in Section 3 of that paper.

\begin{figure}[]
  \begin{center}
    \includegraphics[width=1\textwidth]{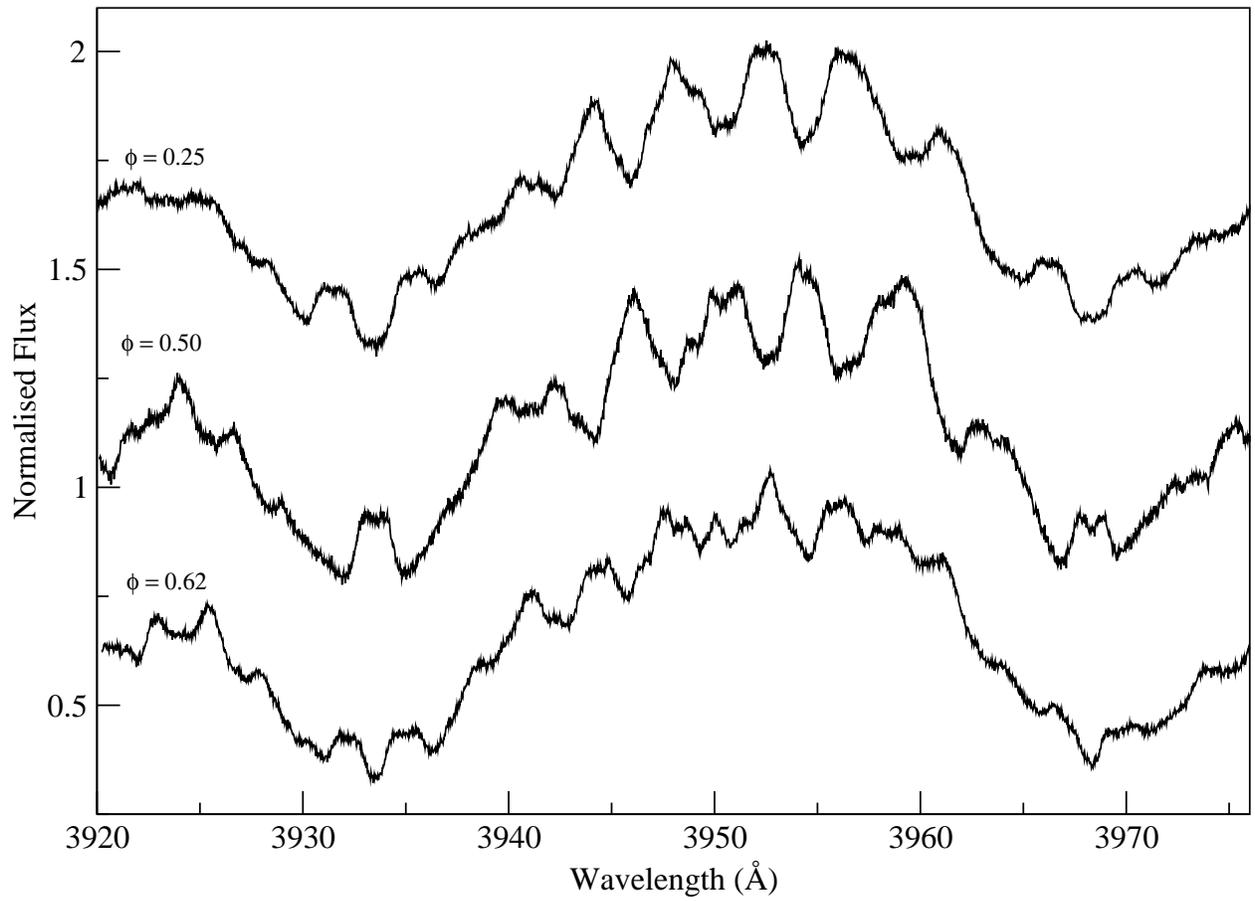}
    \caption[ER Vul Spectra at $\phi$ = 0.25, 0.50, 0.62]{~~~ER Vul spectra normalized to the peak intensity at three orbital phases.\label{ervul_specs}}
  \end{center}
\end{figure}

Of the three nights awarded, two were clear (2002 August 26 and 27), giving us
almost complete phase coverage, $\phi$ = 0.09 $-$ 0.96.
Table~\ref{ervul_data} lists the heliocentric Julian date (HJD) and orbital phase of each exposure.
Harmanec et al.~(2004) supplied an updated ephemeris from photometric
monitoring done at two observatories in Hvar, Croatia and Victoria, Canada. The time of primary eclipse ($\phi$ = 0),
occurred at HJD = 2440182.25628(46)\footnote{Values in brackets are 1$\sigma$
errors.} with a period of 0.698095113(29) d.
Specimen ER~Vul spectra are shown in Figure~\ref{ervul_specs} at the three
phases $\phi$ = 0.25, 0.50 and 0.62.

\section{The Broadening Functions \label{BFs}}

For a close binary with rapidly rotating components such as those of ER~Vul, the
Broadening Function (BF) approach is optimal to obtain radial velocity information for one-dimensional,
radial-velocity domain images of the stellar surfaces. The broadening function approach was used for the first time in
1992 (Rucinski 1992) and subsequently, for analysis
of radial velocities of close binary stars. The technical summaries are
in Rucinski (1999) and Rucinski (2002). The BF method uses a spectrum of a chromospherically inactive
star as a template. The template could either be a synthetic spectrum or a
real spectrum of a sharp-lined (slowly rotating) star that is artificially
Doppler broadened by the BFs.
Subtracting a broadened template spectrum from that of an active binary leaves only
emissions from stellar active regions such as plages or prominences.

The principal properties of the approach, described in detail
in Rucinski (1992, 1999, 2002)\footnote{See also
http://www.astro.utoronto.ca/$\sim$rucinski.}, are based
on the fact that the observed spectrum of a rapidly-rotating
or revolving star can be approximated by a convolution of a
sharp-line star spectrum with a kernel describing the observed
radial velocity field. Thus,
the broadening function $B$, in $P = S \ast B$, is a kernel in
$P(\lambda ') = \int B(\lambda ' - \lambda) \: S(\lambda) \: d \lambda $.
In this integral equation,
$P$ is the ``program'' spectrum and $S$ is the
``sharp'' or the template spectrum. The equation
can be solved by representing the convolution operation
by a set of over-determined algebraic equations; $S$ \& $P$ are usually
defined over thousands of pixels, while $B$ is usually defined
over hundreds of pixels, so the equations can be cast as
a standard least-squares problem with an over-determinacy typically
of $>10$ times.

The use of a radial velocity standard star of similar spectral
type for the template $S$ straightforwardly relates the results
to the velocity scale and insures the use of all, even the weakest, lines
which would not appear in model spectra; this property is
similar to that of the Cross-Correlation Function (CCF). However, the resolution
of the BF depends on and is entirely defined by the
instrumental setup, which is a great advantage over the
CCF which combines all sources of broadening. This
frequently leads to the necessity of excising wide hydrogen lines,
molecular features, etc.
The BF is derived through a linear process, so
that its features can be used directly for estimates of the
the relative light intensities.  A very similar technique, called the Least Squares
Deconvolution (LSD), was developed mainly for magnetic Zeeman
pattern extraction by Donati et al.~(1997) and then used for radial
velocity and line broadening applications (e.g.~Collier Cameron et al.~2002, Reiners \& Schmitt 2003). Both techniques use the Singular
Value Decomposition for determination of the spectral line locations and
broadening, but the LSD is typically used with model spectra while the
BF formalism relies on stellar templates for direct tie-ins to the radial
velocities of standard stars and accounts for weak lines not present
in model spectra.

For
a close binary like ER~Vul, the BF is essentially a projection of the
system into the radial velocity domain. If stars are spherical
and rotate rigidly, the BF is expected to have the
ellipsoidal shape predicted by the Struve
profile (Solana \& Frenley 1997). Spots will appear
as notches in the BF at certain velocities. It is important to recognized that
any emission components will appear as dark notches,
which is perhaps somewhat less intuitive; however, we do not
expect any emission components in the photospheric lines used to determine the BFs in this study.

We used the average spectrum of HD~209458, formed from three spectra
taken a few weeks prior to our
2002 August observing run, as the template (standard)
star with which to calculate the BFs for the ER~Vul spectra.
HD~209458 is a G0~V star making it a good match
to both of ER~Vul's components. Computations of the BFs involve
inversion of large matrices. The computing time can be reduced
if the spectra are reduced in length by re-sampling
to lower resolution. The
full resolution of 0.013 \AA\ per pixel (0.95 km~s$^{-1}$) was
deemed unnecessary; we reduced it two-fold to
the velocity step of 1.9 km~s$^{-1}$.
Considering that the spectral features of ER~Vul
are very broad, there is no significant loss of information from this
step.  The total number of useful pixels in the equal-velocity-step
spectra was 2250 (at 1.9 km~s$^{-1}$), while the wide range
of the ER~Vul velocities dictated a relatively large spectral
window of 700 km~s$^{-1}$ or 369 pixels. Thus, the overall
over-determinacy in the BF solutions was moderate, 6 times,
which was adequate to provide high-quality BF profiles.

Only the upper 2/3 of the ER Vul spectrum (as measured from the peak flux of the pseudo-continuum from in between the two strong Ca II lines) was used in the calculations of the BFs in order to exclude the Ca~II reversals and to eliminate
any departures from the photospheric component
due to chromospheric activity. Besides, only the well defined upper parts of the spectra found
in regions between and outside the
Ca~II cores, which are rich in narrow lines, could provide a good basis for the BF determination.

The most difficult and unusual step in the BF determination
was the normalization of the strongly curved spectrum outside the
Ca~II features. The spectra of ER~Vul are severely blended
and there are no places
where we can see the photospheric
continuum. The normalization was done by an interactive determination of the upper envelope to the
spectrum and division of the spectra by such an approximate
continuum. That this is not the actual photospheric continuum is clearly
indicated by the sums of the BFs which, in spite of the
spectral type of the template (G0V) matching very well with the
of the components of ER~Vul, are not close to unity for each
component, but are instead at the level of $\simeq 0.35$. This means that about
2/3 of the total amount of absorption in the spectral lines of
ER~Vul is located in unaccounted-for line blends. The combined effect
of the blending
is the lowering of the effective continuum and shallower depths of
those lines that remain visible. Because this is still
a linear transformation and
the broadening functions are determined through a
linear process, the total intensities of the BFs are affected while the shapes and positions of the BFs remain
entirely unchanged. As a final step, we normalized the total intensities to 1.

\begin{figure}[]
  \begin{center}
    \includegraphics[angle=270,width=1\textwidth]{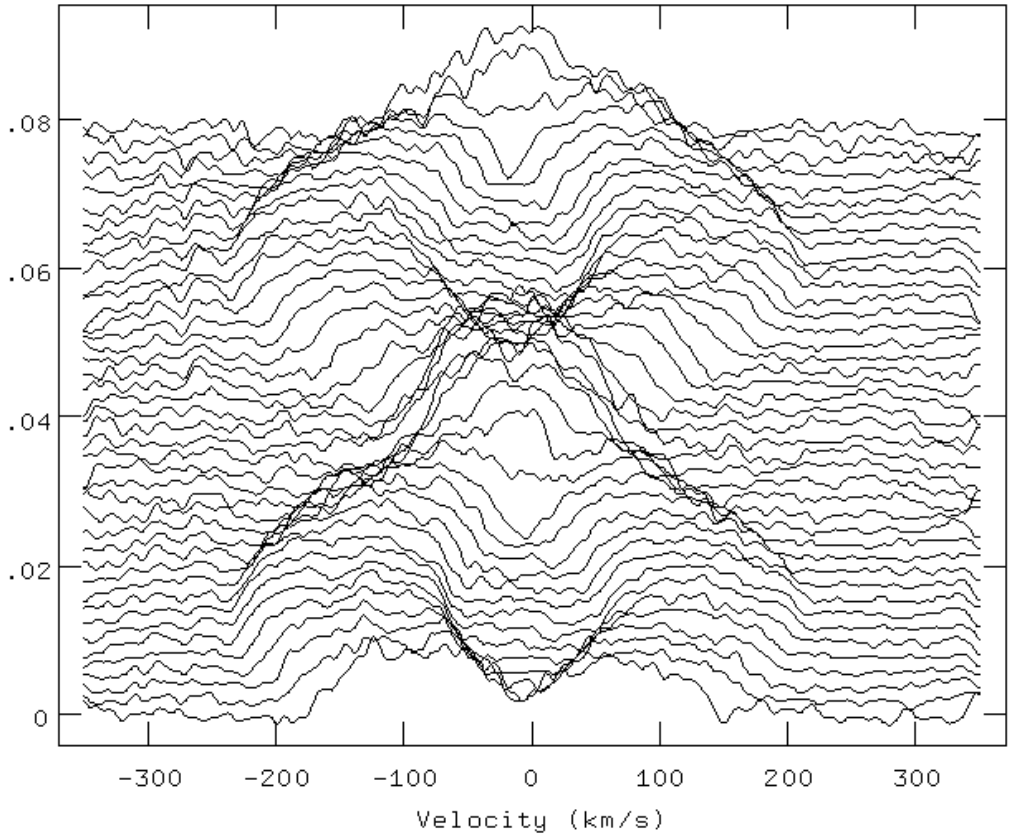}
    \caption[The Broadening Functions]{~~~The broadening functions of ER Vul for all data starting with $\phi$ = 0.09 (bottom curve) to 0.96 (top curve) as a function of BF intensity. The intensities are normalized but are displaced vertically.\label{BF_specplot}}
  \end{center}
\end{figure}

\begin{figure}[]
  \begin{center}
    \includegraphics[width=0.8\textwidth]{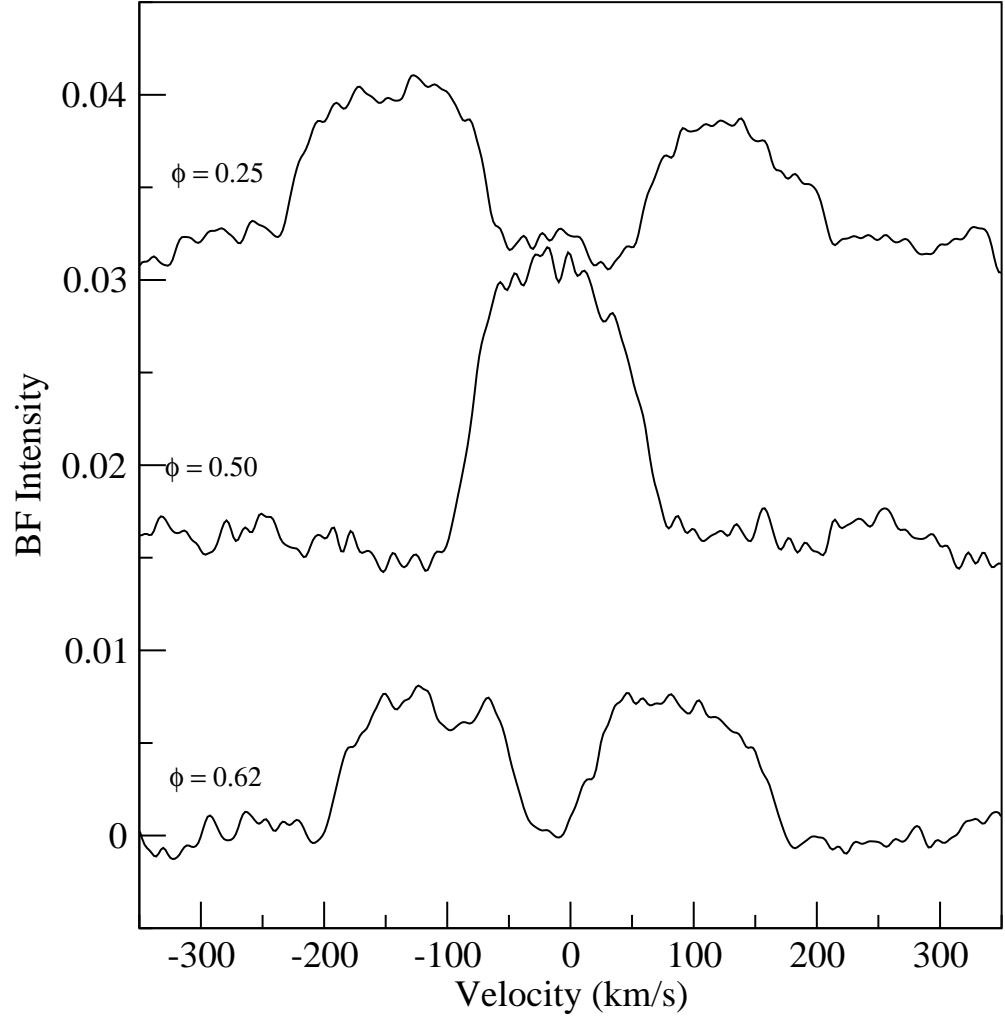}
    \caption[BFs at $\phi$ = 0.25, 0.50, 0.62]{~~~Examples of broadening functions at three orbital phases.\label{BF_3phases}}
  \end{center}
\end{figure}

The final BFs are very well defined and clearly show both
of the components. All determined BFs are plotted in
Figure~\ref{BF_specplot}.
Figure~\ref{BF_3phases} provides a closer look at
the BFs at $\phi$ = 0.25, 0.50 and 0.62. We can determine RVs and hence the system orbital parameters from the BFs as measured in the next section. We do not see any evidence of bright or dark spots on the stellar surfaces which would appear as travelling notches in the BFs implying that more than 50\% of the stellar disk may be covered by one or more spots. This is consistent with Sokoloff \& Piskunov (2002) who observe a large, high-latitude cool spots on ER Vul.

\clearpage

\begin{deluxetable}{ccrrc}
\tabletypesize{\footnotesize}
\tablecaption{ER~Vul: 2002 August Observations\label{ervul_data}}
\tablewidth{0pt}
\tablehead{
\colhead{HJD $-$ } &
\colhead{Orbital} &
\colhead{RV (P)\tablenotemark{a}} &
\colhead{RV (S)\tablenotemark{a}} &
\\
\colhead{2452000} &
\colhead{Phase} &
\colhead{km s$^{-1}$} &
\colhead{km s$^{-1}$} &

}
\startdata
512.773	&	0.090	&	-86.0	&	65.4	\\
512.789	&	0.113	&	-99.3	&	82.9	\\
512.806	&	0.137	&	-113.5	&	92.1	\\
512.822	&	0.160	&	-124.5	&	101.5	\\
512.839	&	0.184	&	-132.7	&	115.5	\\
512.855	&	0.207	&	-138.3	&	121.7	\\
512.871	&	0.230   &	-141.2	&	126.0	\\
512.887	&	0.253	&	-143.6	&	125.4	\\
512.903	&	0.276	&	-145.1	&	126.9	\\
512.919	&	0.299	&	-137.1	&	119.9	\\
512.935	&	0.321	&	-130.2	&	116.3	\\
512.950	&	0.344	&	-121.7	&	106.6	\\
512.967	&	0.367	&	-112.8	&	93.1	\\
512.983	&	0.390	&	-102.1	&	82.5	\\
512.999	&	0.414	&	-71.8	&	73.9	\\
513.015	&	0.437	&	-21.3	&	60.6	\\
513.031	&	0.460	&	-14.1	&	-14.1	\\
513.047	&	0.482	&	-14.8	&	-14.6	\\
513.063	&	0.505	&	-13.7	&	-13.7	\\
513.734	&	0.467	&	-16.8	&	-16.4	\\
513.750	&	0.489	&	-15.0	&	-15.4	\\
513.766	&	0.512	&	-13.3	&	-14.4	\\
513.782	&	0.535	&	-7.6	&	-8.1	\\
513.798	&	0.558	&	0.4	&	0.7	\\
513.814	&	0.581	&	39.7	&	-79.1	\\
513.829	&	0.603	&	71.3	&	-100.3	\\
513.845	&	0.626	&	81.9	&	-116.3	\\
513.861	&	0.648	&	92.3	&	-128.3	\\
513.877	&	0.671	&	103.3	&	-137.9	\\
513.893	&	0.694	&	109.1	&	-146.3	\\
513.909	&	0.717	&	117.1	&	-149.5	\\
513.924	&	0.739	&	112.5	&	-148.5	\\
513.940	&	0.761	&	114.2	&	-154.6	\\
513.955	&	0.784	&	112.7	&	-147.4	\\
513.971	&	0.806	&	104.9	&	-142.6	\\
513.987	&	0.829	&	101.6	&	-134.2	\\
514.003	&	0.852	&	88.3	&	-122.2	\\
514.019	&	0.875	&	77.6	&	-108.4	\\
514.035	&	0.898	&	68.4	&	-94.6	\\
514.051	&	0.920	&	43.4	&	-68.1	\\
514.066	&	0.942	&	1.9	&	2.0	\\
514.082	&	0.965	&	-8.5	&	-8.8	\\

\enddata
\tablenotetext{a}{RV measured from the centroid of the BF.}

\end{deluxetable}
\clearpage

\subsection{Radial Velocities from the BFs}

\begin{figure}[]
  \begin{center}
    \includegraphics[width=1\textwidth]{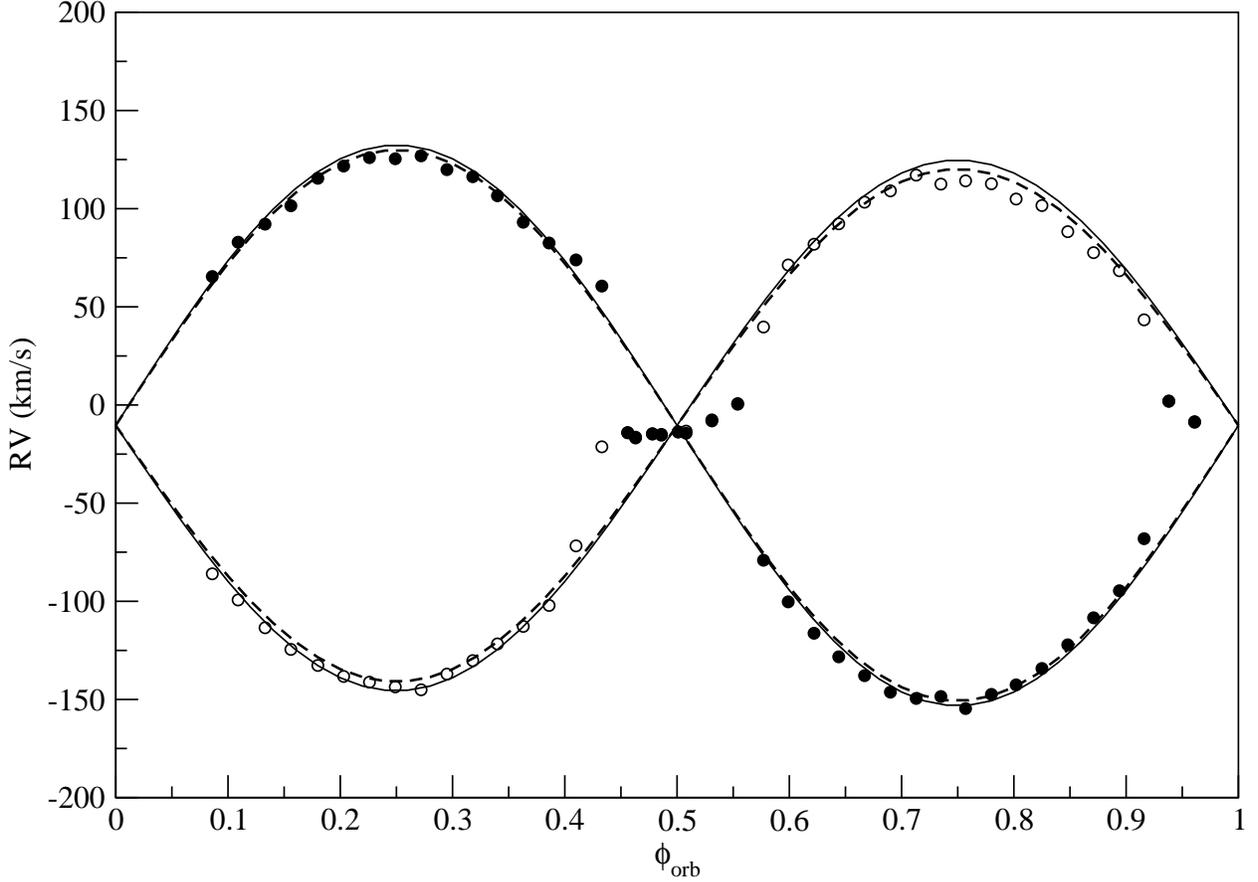}
    \caption[Radial Velocity of ER Vul]{~~~The ER Vul  radial velocities determined from the BFs as a function of orbital phase. The RVs of the primary are represented by the open circles, and the secondary by the filled circles. The data are listed in Table~\ref{ervul_data}.  The solid curves are orbital fits using values from Duemmler et al.~(2003) and the dashed curves are the least-squares best fits to our data.  See text for more details.  The uncertainties in the RVs is $\approx$ 0.5 km~s$^{-1}$ for the values outside of the eclipses. During the eclipses, the BFs of the components are blended making the RVs difficult to measure. The error in $\phi$ is $<$~0.0001 since the ephemeris is very well known. \label{ervul_RV}}
  \end{center}
\end{figure}

Even though RV curves for the ER Vul system have been measured repeatedly, we demonstrate the quality and extent of information one can extract from the BFs. Despite the severe line blending and the limited spectral range, we were able to determine good RV curves from the BFs. We measured radial velocities (RVs) for each component of ER Vul by fitting Gaussian curves to the BFs of each star at each phase. The velocity at the line centroid was measured with an uncertainty of 0.5 km~s$^{-1}$. The RV curve as a function of phase is plotted in Figure~\ref{ervul_RV}. The RVs  near conjunction (0.9 $\leq \phi \leq$ 0.1 and 0.4 $\leq \phi \leq$ 0.6) are inaccurate because only one Gaussian could be fitted to the blended BFs. The solid curves are sine functions using the system parameters measured by Duemmler et al.~(2003). The velocity amplitudes of the primary and secondary components are $K_{P}$ = 135.20 $\pm$ 0.63 km~s$^{-1}$  and $K_{S}$ = 142.82 $\pm$ 0.76~km~s$^{-1}$, the eccentricity $e$ = 0 and the systemic velocity $\gamma$ = 25.49 $\pm$ 0.39~km~s$^{-1}$.  The dashed curves are the best-fitting sine functions to the data using the RVs outside of the conjunction region. For these, $K_{P}$ = 130.55 km~s$^{-1}$, $K_{S}$ = 140.25 km~s$^{-1}$, $e$ = 0 and $\gamma$ = 26.68~km~s$^{-1}$.  The RMS scatter of the data relative to the solid curves are 6.15 km~s$^{-1}$ for the primary and 4.30 km~s$^{-1}$ for the secondary.  The deviation of the primary by $\approx$~$-$10 km~s$^{-1}$  at $\phi \approx$ 0.75 may be due to spectral line distortions caused by less stellar activity relative to the rest of the star and could be an indication of relatively {\it inactive} longitudes.  The RV of the secondary also deviates at its maximum, but only by $\approx$~$-$5 km~s$^{-1}$.

\section{Ca II Emission in the ER Vul System \label{CaII}}

\subsection{Extracting the K Emission}

\begin{figure}[]
  \begin{center}
    \includegraphics[width=1\textwidth]{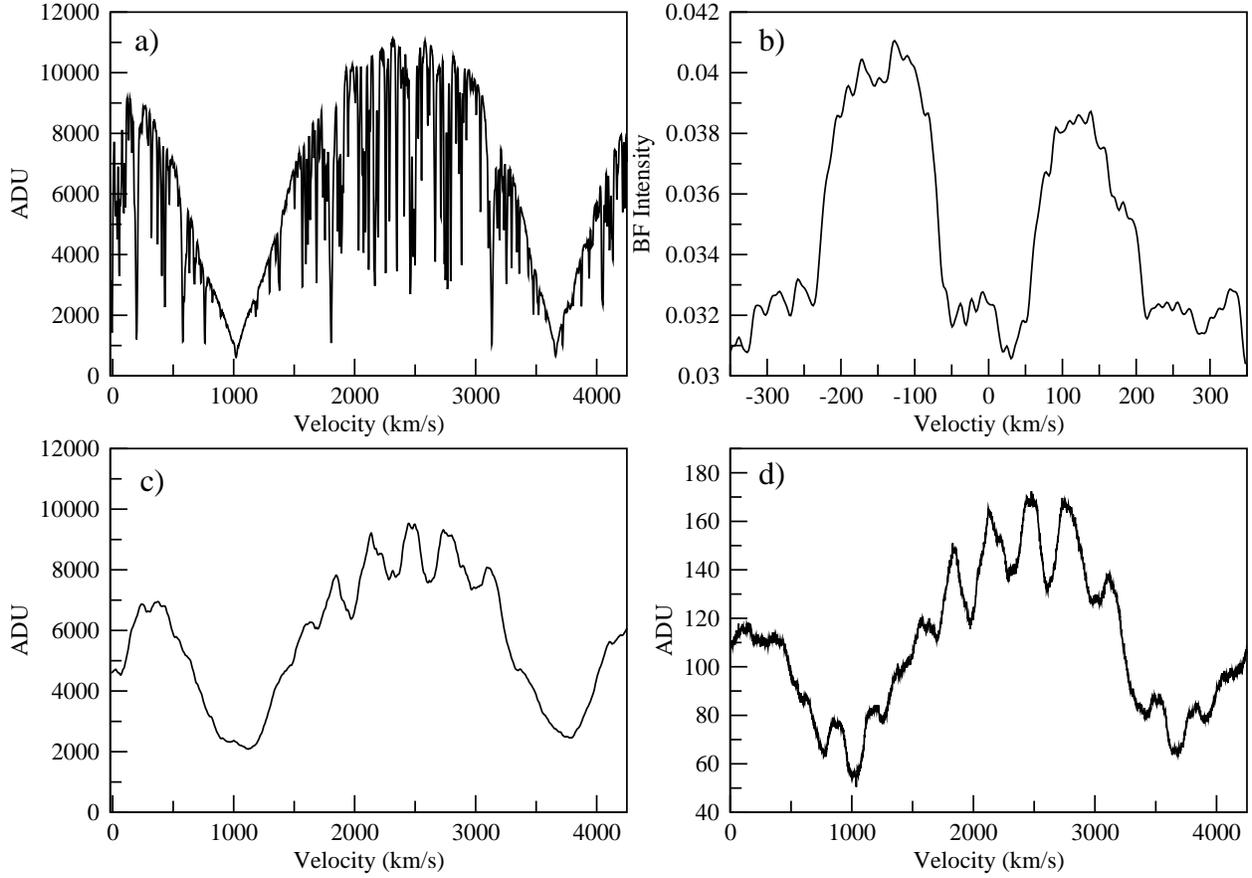}
    \caption[The Convolution]{~~~a) The template spectrum of the sharp-lined star HD 209458. b) The BF calculated for $\phi$ = 0.25. c) The convolution of the template spectrum with the BF. d) The ER Vul spectrum at $\phi$ = 0.25.  \label{conv}}
  \end{center}
\end{figure}

The rapid rotation and orbital velocities of the two components of ER Vul introduce severe Doppler broadening and line blending making it difficult to distinguish the contribution from each star separately.  The BF technique allows the chromospheric emission for each star to be isolated from the rest of the stellar flux by subtracting a template spectrum convolved with the appropriate BF. Figure~\ref{conv} illustrates the steps involved at a given phase: (a) the sharp-lined spectrum of HD~209458 that we used as our standard, (b) the BF measured from the ER Vul spectrum at $\phi$ = 0.25 (quadrature), (c) the convolution of the BF with the sharp-lined template, and (d) the observed ER Vul spectrum at $\phi$ = 0.25.

\begin{figure}[]
  \begin{center}
    \includegraphics[width=1\textwidth]{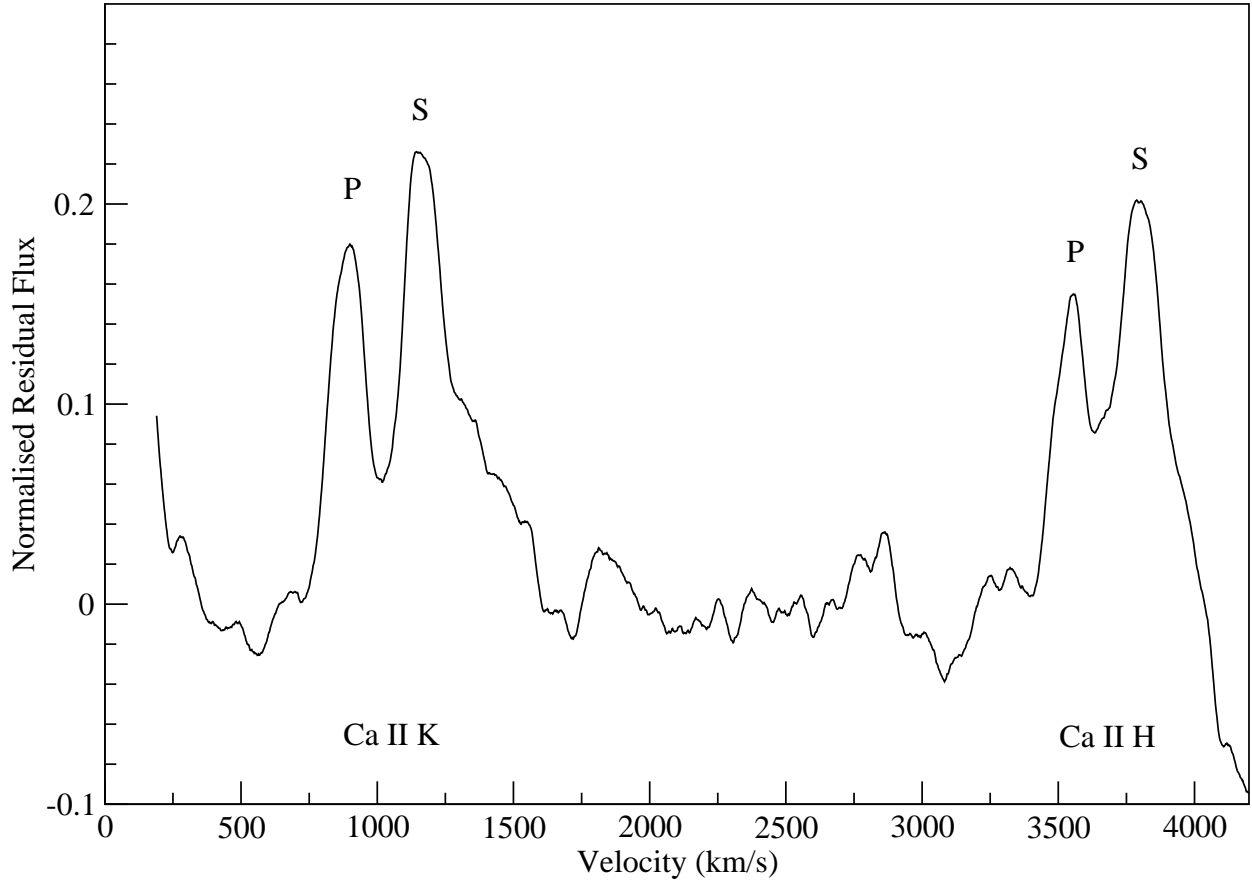}
    \caption[H \& K Emission at $\phi$ = 0.25.]{~~~The full Ca II H \& K residual spectrum of the broadened template subtracted from the corresponding ER Vul spectrum at $\phi$ = 0.25. The emission from the primary and secondary are labeled `P' and `S', respectively.\label{diff10}}
  \end{center}
\end{figure}

This procedure to remove the photospheric contribution to the spectrum works well as is seen in the residual spectrum (with no normalization) in Figure~\ref{diff10}. The three origins of H \& K emission remain: the chromospheres of the primary and secondary stars, and a broad emission which, at this phase ($\phi$ = 0.25), is highly redshifted. Unfortunately, the 60-\AA-wide spectrum cuts off the broad emission red-ward of the H line so we will not discuss the H emission much further as it suffers from truncation in the convolution. The fluctuations seen in the figure between the H and K lines (1800 $-$ 3200 km~s$^{-1}$) are most likely due to two processing issues: 1) the spectral type of HD~209458 matches ER Vul's primary star very well, but differs by 3 sub-classes for the secondary; and 2) when comparing broad-lined spectra with highly-blended features, the normalization becomes a tricky task. This challenge arises during the calculation of the BFs and then again when setting the pseudo-continuum level of the standard and stellar spectra before subtraction. The stellar and template spectra used to produce Figure~\ref{diff10}, were merely set to the same maximum level. The larger fluctuations in this region may also be attributed to the presence of the large, high-latitude cool spots observed by Sokoloff \& Piskunov (2002).  Such spots would create flat-bottomed photospheric lines when highly rotationally broadened causing small bumps when the broadened template is subtracted.

Before subtracting the convolved templates, the ER Vul spectra had to be rebinned to the same 1.9 km~s$^{-1}$ per pixel as the BFs and templates. A cross-correlation was then performed with IRAF's {\it fxcor} routine of each ER Vul spectrum with its corresponding template, using only the photospheric flux between the H and K absorption lines gives the best alignment of the spectra before subtraction. The average measurement error in the velocity shift was 4.5 km~s$^{-1}$ (or 2.4 pixels).

\begin{figure}[]
  \begin{center}
    \includegraphics[angle=270,width=1\textwidth]{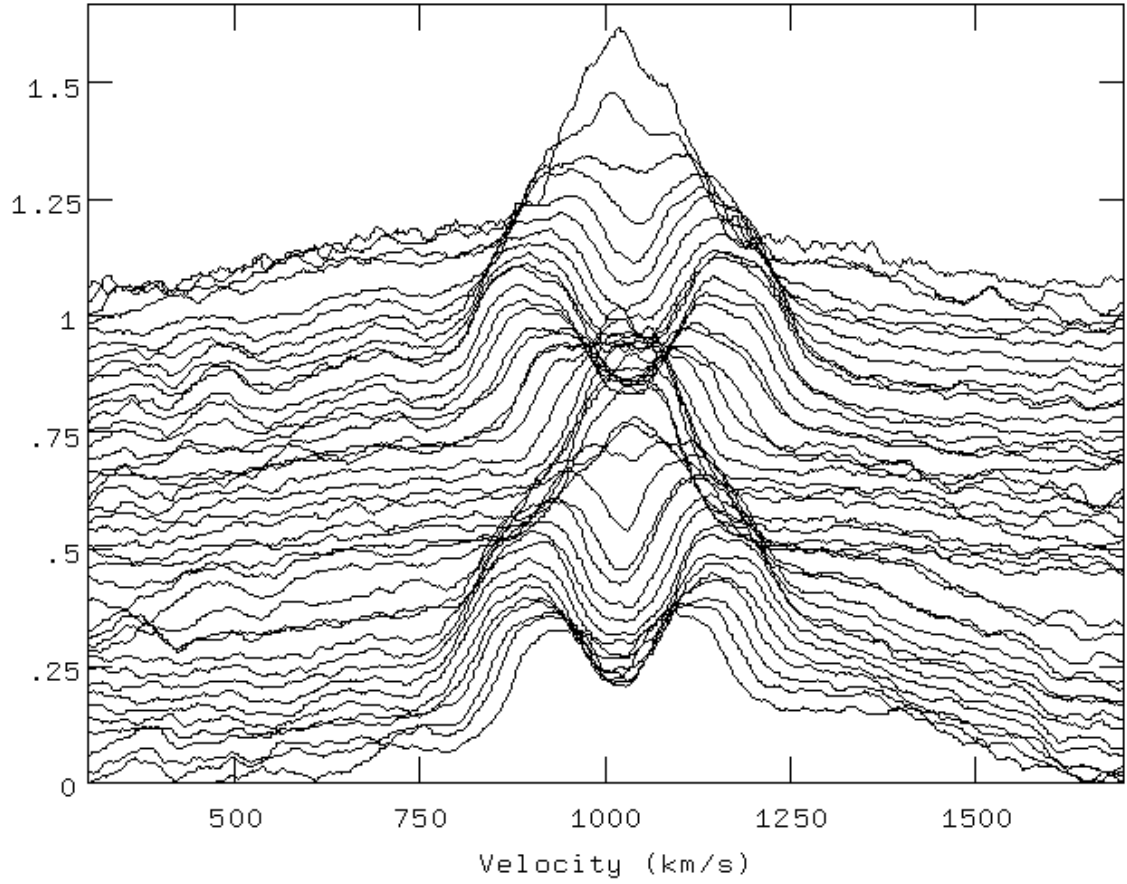}
    \caption[The Ca II K Emission for ER Vul]{~~~The Ca II K emission of ER Vul (smoothed by 9 pixels) at all observed phases starting at $\phi$ = 0.09 (bottom curve) to 0.96 (top curve).  The vertical axis is normalized residual flux with arbitrary vertical shift.  \label{Kdiffs_specplot}}
  \end{center}
\end{figure}

\begin{figure}[]
  \begin{center}
    \includegraphics[width=1\textwidth]{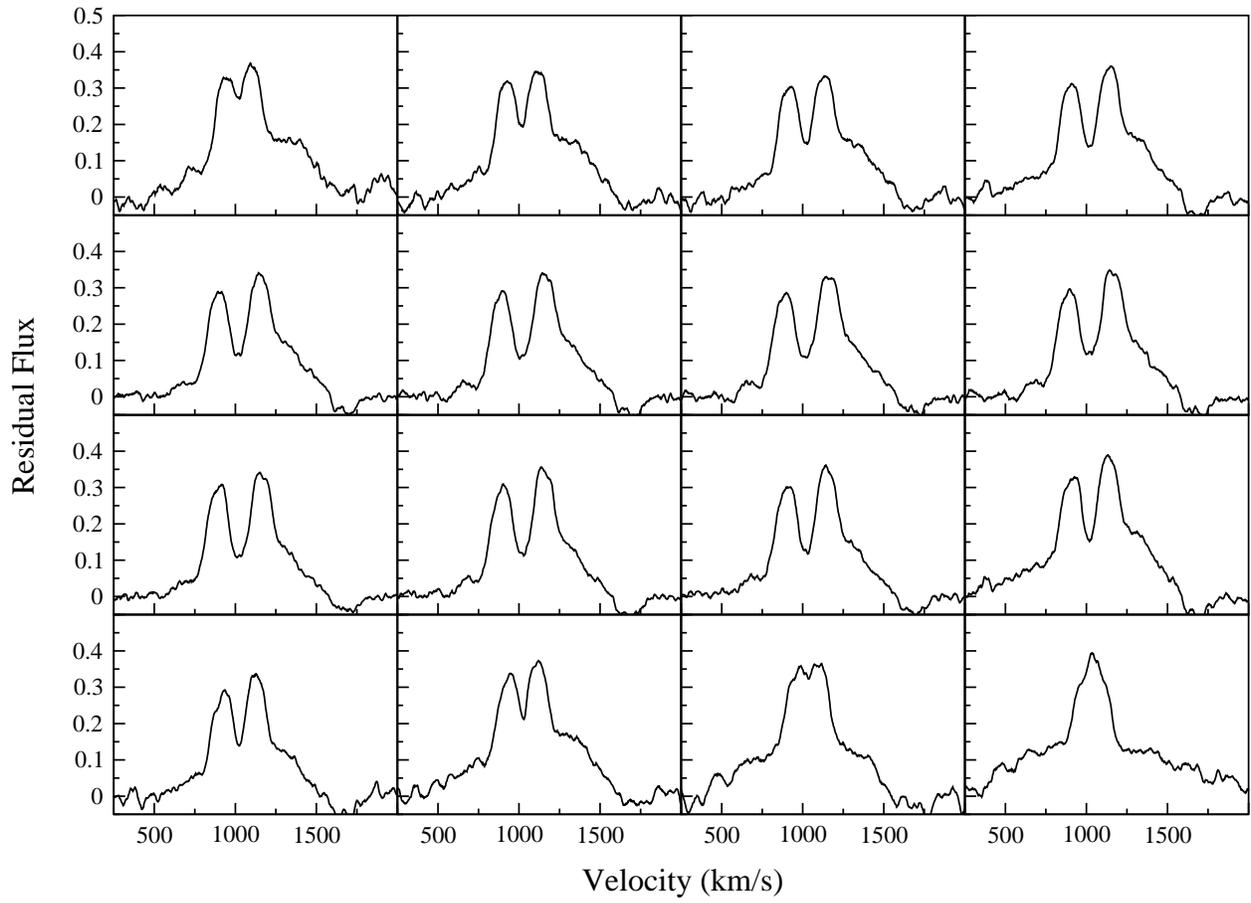}
    \caption{~~~Ca II K residual emission for phases ranging from 0.090 to 0.437.\label{diffs1}}
  \end{center}
\end{figure}

\begin{figure}[]
  \begin{center}
    \includegraphics[width=1\textwidth]{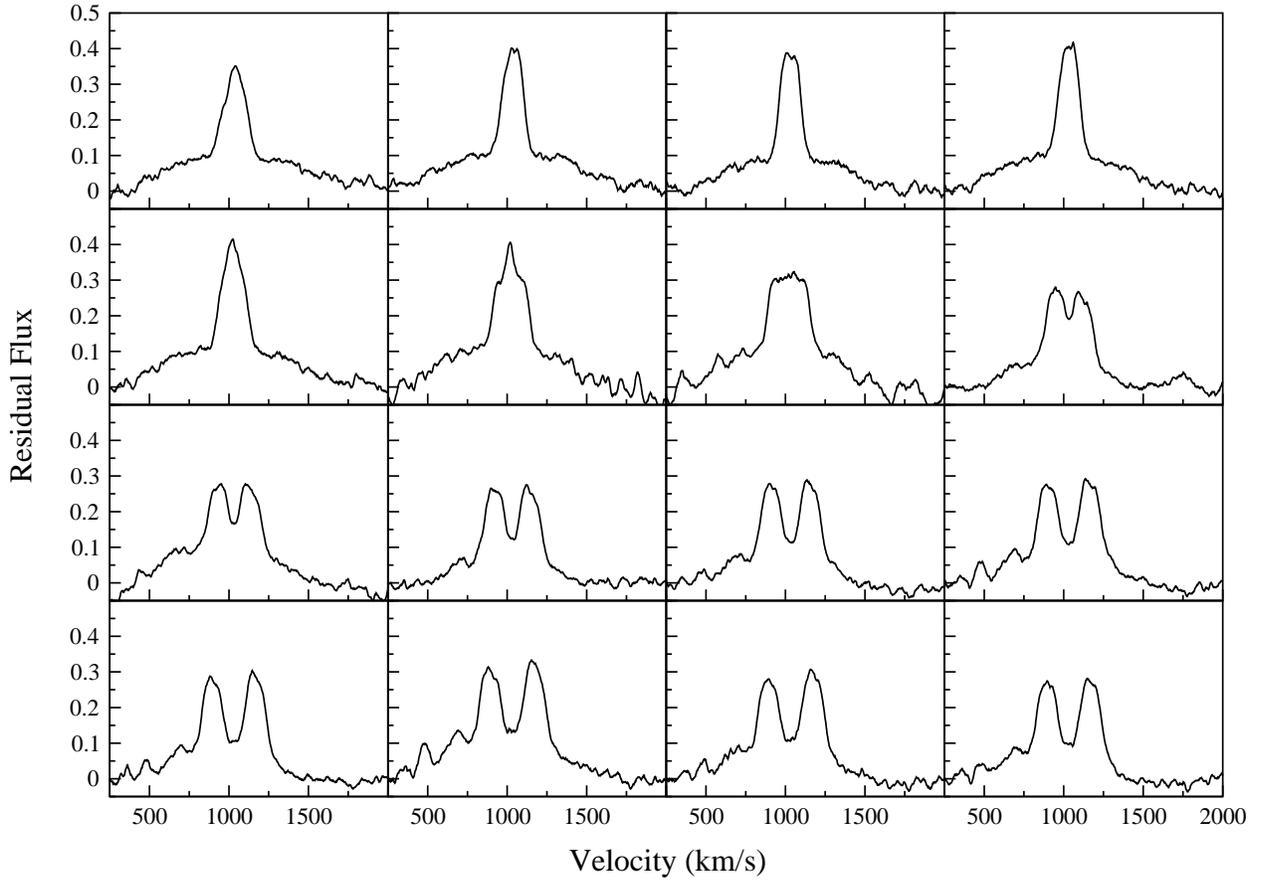}
    \caption{~~~Ca II K residual emission for phases ranging from 0.460 to 0.761.\label{diffs2}}
  \end{center}
\end{figure}

\begin{figure}[]
  \begin{center}
    \includegraphics[width=1\textwidth]{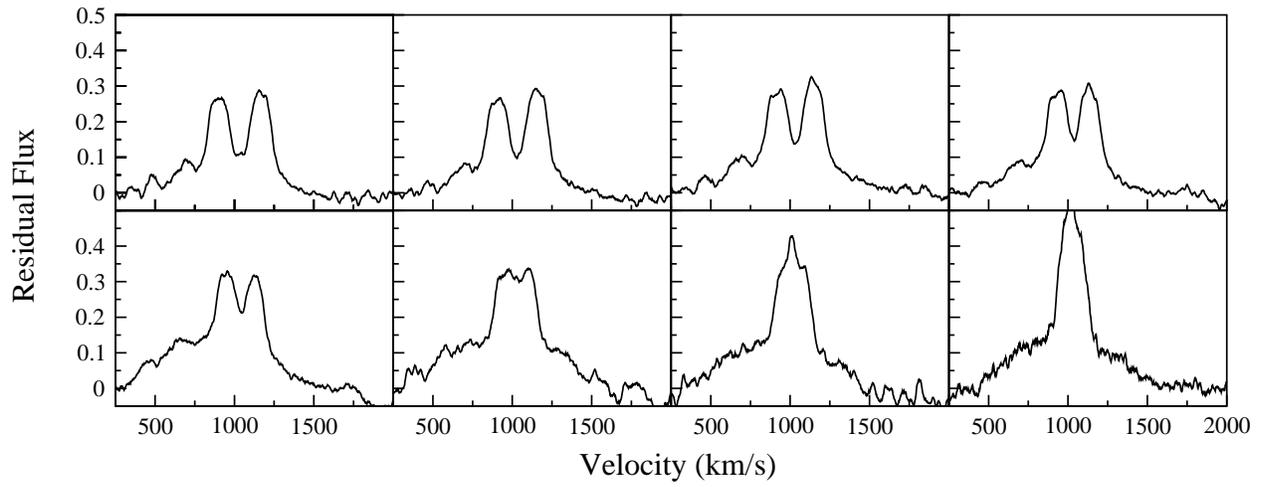}
    \caption{~~~Ca II K residual emission for phases ranging from 0.781 to 0.965. \label{diffs3}}
  \end{center}
\end{figure}

Both the broadened template and the ER Vul spectrum were normalized by fitting a low-order polynomial to the regions outside the H and K cores before subtraction. The templates were subtracted from the corresponding ER Vul spectra to isolate the chromospheric contribution. The residuals for all K lines are plotted in Figure~\ref{Kdiffs_specplot} starting with $\phi$ = 0.09 at the bottom and individually plotted in Figures~\ref{diffs1}, \ref{diffs2} and \ref{diffs3}.

\subsection{Origins of Emission}

\begin{figure}[]
  \begin{center}
    \includegraphics[width=1\textwidth]{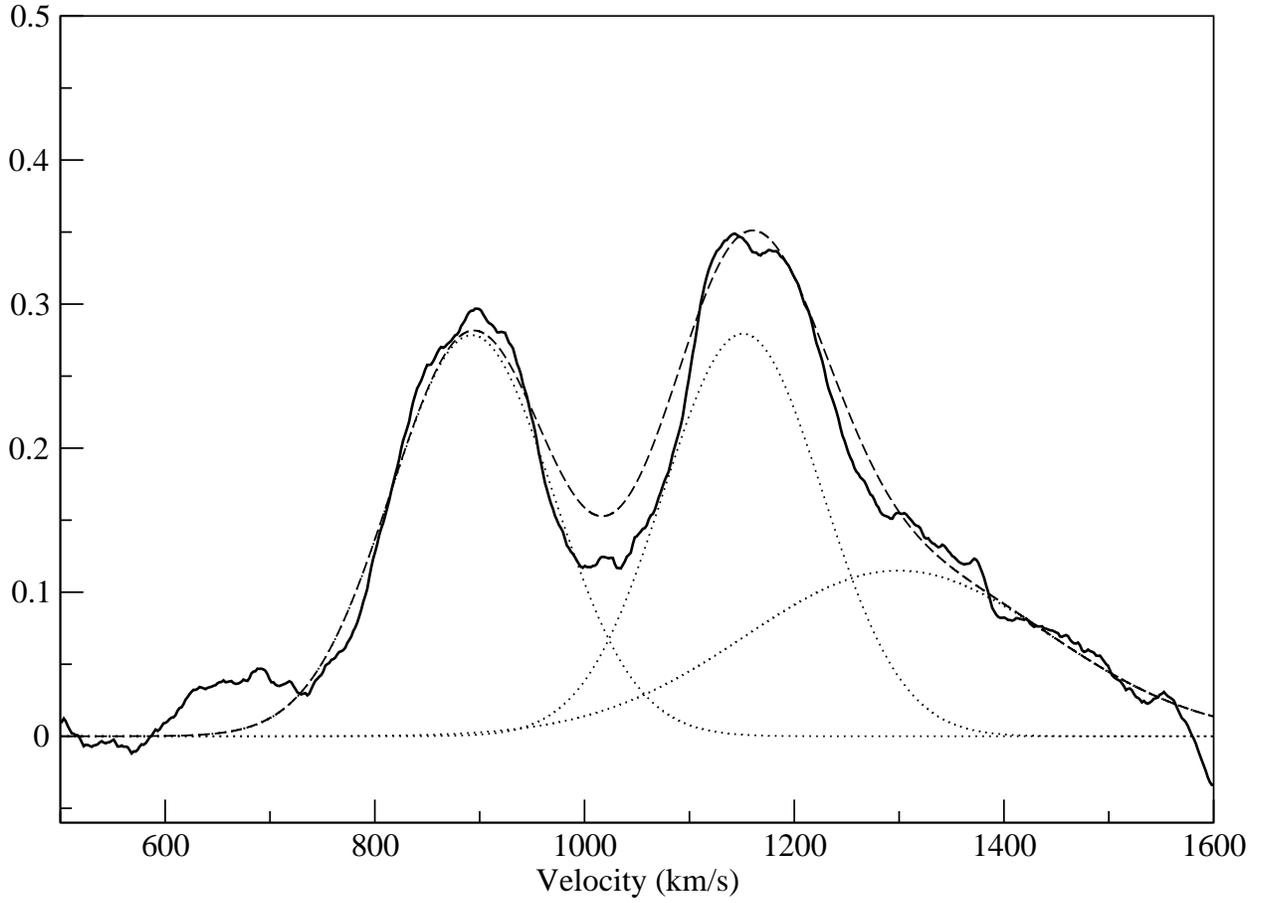}
    \caption{~~~The K emission for $\phi$ = 0.25 (solid curve) deblended into three Gaussians (dotted curves).  The dashed curve is the sum of the three Gaussians.\label{gaussians}}
  \end{center}
\end{figure}

For phases outside of conjunction,
the residual K emission is well fitted with three Gaussian profiles indicating three separate origins. An example is shown in Figure~\ref{gaussians}. This is a necessary step to isolate the contributions made by each.
The two narrower peaks are clearly the Ca II K emission from each of the stars. Neither the primary nor secondary star's K emission strength appears to vary much throughout the orbit. Through their UV and X-ray observations, Vilhu \& Rucinski (1983) suggest that there is a level at which the emission from RS CVn systems becomes saturated. Gunn \& Doyle (1997) base their assumption that ER Vul's chromospheric emission is close to saturation on the very large filling factor they obtained. Saturation of the chromospheric emission may be another explanation for the lack of variation seen the K emission of the two stars. The extremely strong emission in practically every spectral activity indicator supports this possibility.

\begin{figure}[]
 \begin{center}
    \includegraphics[width=0.7\textwidth]{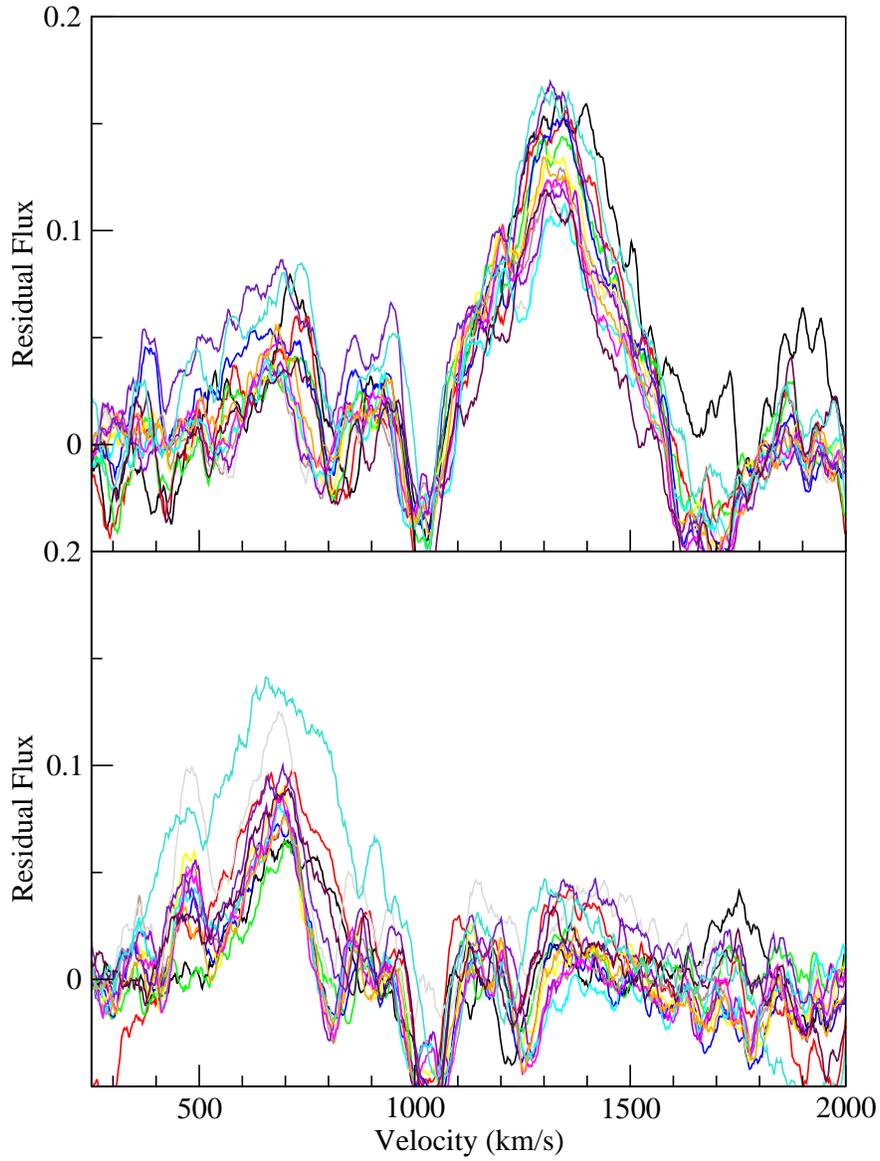}
    \caption{~~~The broad emission after the K emission from the two stars has been subtracted overlaid (excluding phases nearing conjunction); Top: from $\phi$ = 0.090 $-$ 0.39 and Bottom: $\phi$ = 0.603 $-$ 0.898 \label{humps}}
  \end{center}
\end{figure}

The third source is the is highly-redshifted broad emission between $\phi$ = 0.09 and 0.39. It is shifted 320 km~s$^{-1}$ relative to the center velocity of K line. In Figure~\ref{humps} (top), the Gaussian fits for the two stars were removed and the remaining broad emission for each phase in the range is overplotted. The average FWHM of the Gaussian fits to these is 290 km~s$^{-1}$. The peaks do not shift much through this phase range either.
During the opposite orbital phases, $\phi$ = 0.60 $-$ 0.90, the excess emission appears narrower with a FWHM ranging between 130 and 270 km~s$^{-1}$ and with smaller peaks. However, the peaks are again shifted 320 km~s$^{-1}$ from center but now blueward.  The broad emission at each phase within this range is plotted in Figure~\ref{humps} (bottom).

\begin{figure}[]
  \begin{center}
    \includegraphics[width=1\textwidth]{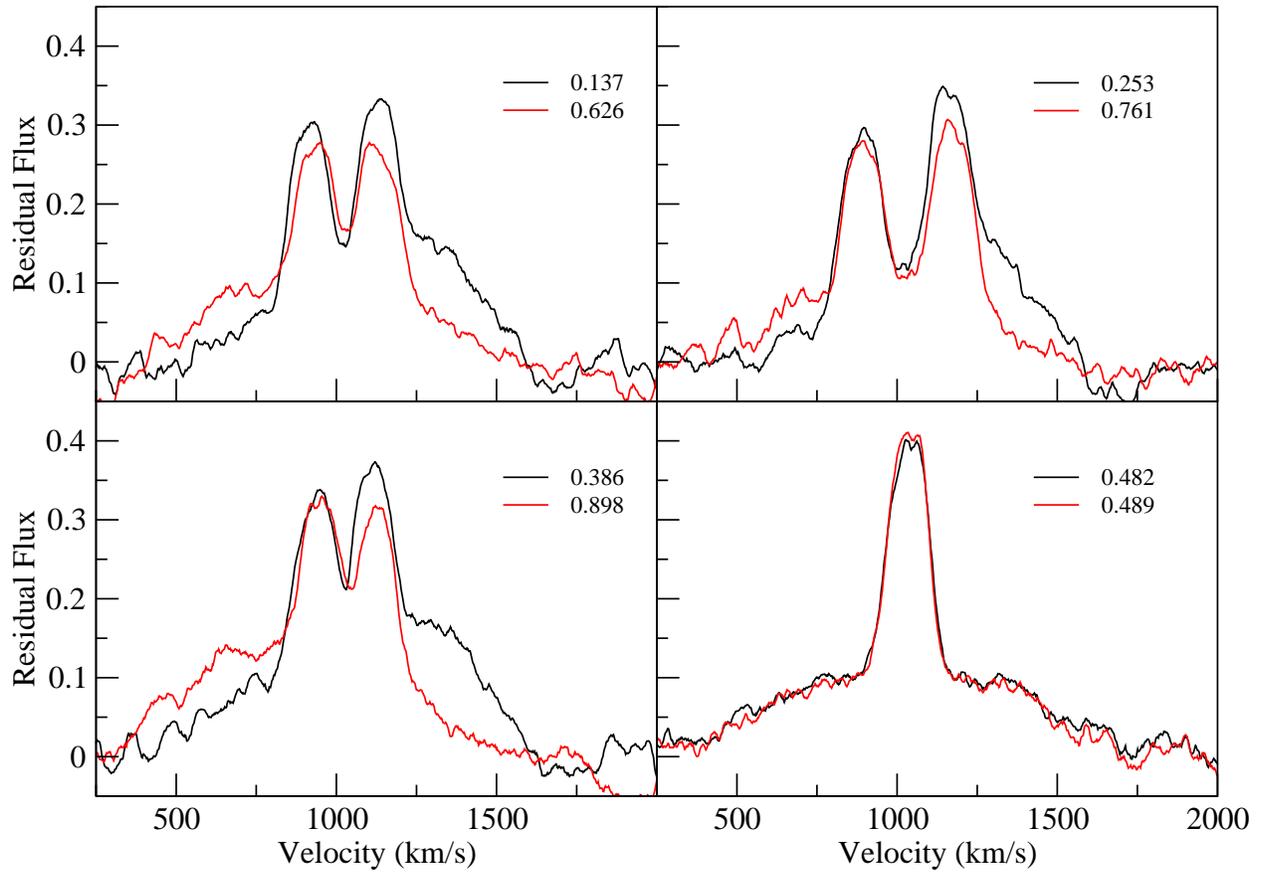}
    \caption{~~~Ca II K residuals at opposite phases, except for the last panel that shows residuals at nearly the same phase but taken on two different nights.\label{opposite_diffs}}
  \end{center}
\end{figure}

The scenario that best fits the observed Ca~II emission from the ER Vul system includes:
1) high-latitude active regions on one or both of the stars that are covering large areas on the stellar surface and
2) a high-velocity stream of hot gas flowing on to the secondary in a widening spiral arriving on the surface at approximately 30 $-$ 60$^\circ$ beyond the sub-binary longitude.  The decreased flux of the stream during $\phi$ = 0.60 $-$ 0.90 is due to its partial eclipse by the secondary. The change in velocity and flux of the broad emission is evident in Figure~\ref{opposite_diffs} where a sample of K residuals are overplotted onto their `opposite phase' residuals.

\begin{figure}[]
  \begin{center}
    \includegraphics[width=1\textwidth]{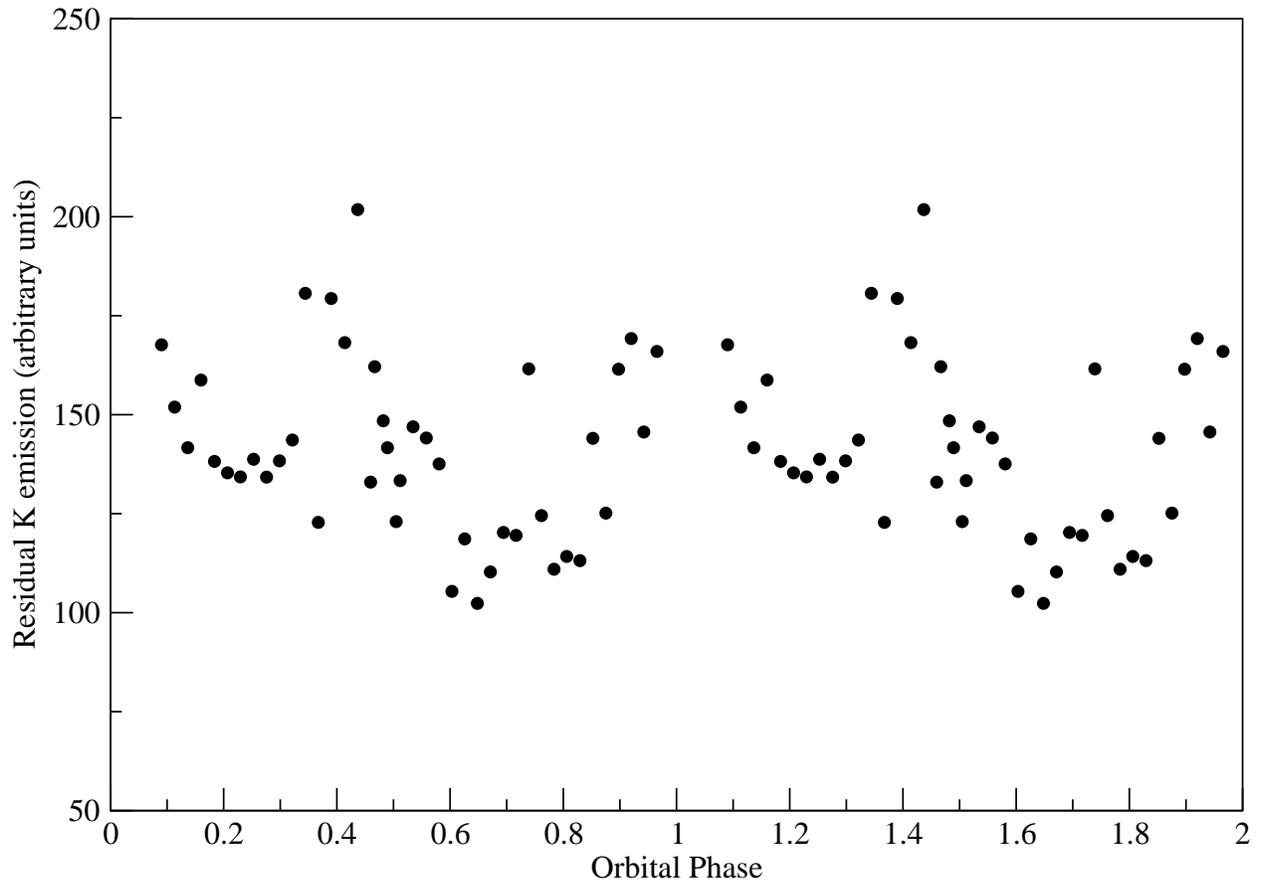}
    \caption{~~~The integrated Ca II K emission of the residuals displayed in Figures~\ref{diffs1}, \ref{diffs2}, \ref{diffs3}.\label{Kemission_phi}}
  \end{center}
\end{figure}

The active region on each star is most likely not confined to a narrow longitudinal range but may cover as much as 50\% of the hemisphere. However, it appears as if the center of the broad emission (and any associated surface hot spot) leads the sub-binary point by $\sim$ 30 $-$ 40$^{\circ}$.
%Even though it is difficult to compare integrated fluxes of the residuals do to inconsistencies in continuum fitting,
Plotting the integrated residual K flux as a function of orbital phase (Figure~\ref{Kemission_phi}),
we see an increase in K emission that peaks near $\phi$ = 0.38 and has a width of about 90 degrees. In this case, the reflection effect as the cause of the heating as suggested by Piskunov (1996) is not supported since the hot spot emitting the excess Ca II is not centered on the sub-binary point.

Our results are consistent with Zeinali et al.~(1995) who also saw evidence of a gaseous stream flowing between the two components in their photometric data of ER Vul, however opposite in direction. More recently, Harmanec et al.~(2004) interpreted phase-locked structure in their analysis of photometric data spanning over 50 years as evidence of gas streams that are projected against the disk of the primary at phases after secondary eclipse. In addition to streams, Ar\'{e}valo et al.~(1988) explained photometric peculiarities in ER Vul as due to circumstellar matter. Hall \& Ramsey (1994) surveyed ten RS CVn binary systems (not including ER Vul) on which they have shown that large regions of extended, prominence-like material are common features with moderate- to high-velocity flows ranging from several tens of km~s$^{-1}$ to 170 km~s$^{-1}$. It is clear that high-velocity streams, prominences, and circumstellar material are all part of ER Vul's dynamic environment.

\begin{figure}[]
  \begin{center}
    \includegraphics[width=0.8\textwidth]{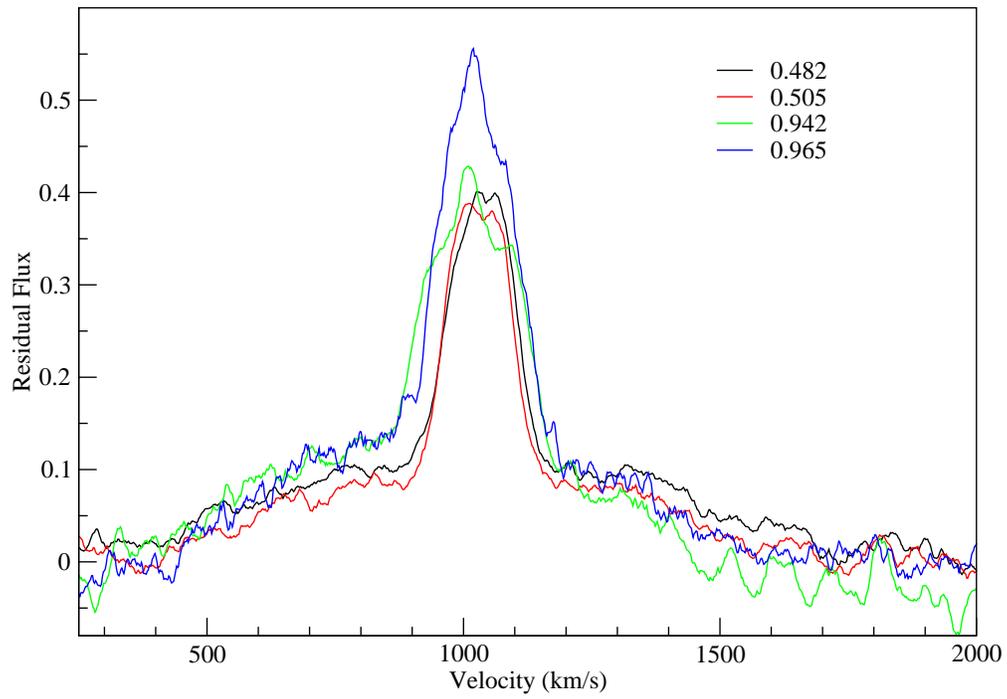}
    \caption{Ca II K residual emission during four phases near conjunction.  Note the effect of a flare at $\phi$ =  0.965. \label{flare}}
  \end{center}
\end{figure}

At the very end of our observations, we caught a rather strong flare at $\phi$ = 0.965 as shown in Figure~\ref{flare}. It contains $\approx$~13\% more power than the emission at the other phases near conjunction. We cannot say from which star the flare originated.

\section{The Photometric Light Curve \label{photometry}}

There is likely to be a spot on the surface of the secondary associated with the impact of the high-velocity stream of hot plasma. This is consistent with observations of a more active secondary as compared to the primary by Gunn \& Doyle (1997) and \c{C}akirli et al.~(2003). A photospheric spot, however, is not detected as a travelling feature in the K residuals or the BFs, nor as a distortion in the RVs measured from the BFs.

Harmanec et al.~(2004) executed concurrent photometric observations of ER Vul in order to search for photospheric indicators of activity. To investigate if a hot spot on the cooler star is consistent
with the light curve of ER Vul, a suite of model light curves were generated
using a recent version of the code described by
Wilson \& Devinney (1971) and Wilson (1979, 1990, 1993).
The models were compared with normal points generated from the
V-band light curve recorded at the same
epoch as the spectroscopic observations.

The geometrical parameters of ER Vul were set first by
considering spot-free models. ER Vul has been the object of a number of
photometric investigations (e.g.~Harmanec et al.~2004; Olah et al.~1994;
Hill et al.~1990), and so the solution computed by
Harmanec et al.~(2004) from combined UBVRI and radial velocity data was used as a
starting point. The temperatures of the two stars were fixed
at the values adopted by Hill et al.~(1990).

The final adopted
inclination and mean stellar radii, computed after making minor
adjustments to match the depths and widths of the eclipses, are summarized
in Table~\ref{model_data}. The ratio of the stellar radii is intermediate between
the values found by Hill et al.~(1990), who found that the components have the same
radii, and Harmanec et al.~(2004), who use r$_2$/r$_1$ = 0.64.

%The final adopted inclination ($i$ = 67.99$^{\circ}$) and mean stellar radii ($r_1$ = 0.291 and $r_2$ = 0.240, in units of orbital separation) were computed after making minor adjustments to match the depths and widths of the eclipses.
%are summarized in Table~\ref{model_data}. The ratio of the stellar radii are intermediate between the values found by Hill et al.~(1990), who found that the components have the same radii, and Harmanec et al.~(2004), where r$_2$/r$_1$ = 0.64. The spot in the model is at a longitude of 36$^{\circ}$ with a radius of 54$^{\circ}$ and a temperature of 1.015 of the secondary star's effective temperature.

\clearpage
\begin{deluxetable}{ccr}
\tabletypesize{\footnotesize}
\tablecaption{Light curve model parameters\label{model_data}}
\tablewidth{0pt}
\tablehead{
\colhead{Parameter} &
\colhead{Value} &
\\
%\colhead{} &
%\colhead{} &
}
\startdata

$i$ & 67.99$^\circ$ \\
$r_1$ & 0.291\tablenotemark{a}\\
$r_2$ & 0.240 \\
spot long & 36$^\circ$ \\
spot rad & 54$^\circ$ \\
spot temp & 1.015T$_{eff,2}$\tablenotemark{b}\\

\enddata
\tablenotetext{a}{in units of orbital separation}
\tablenotetext{b}{the effective temperature of the secondary star}
\end{deluxetable}
\clearpage

Keeping the geometric parameters fixed, a second set of runs were made in which
a hot spot was added to the secondary, with parameters chosen to reproduce the magnitude
difference between the out of eclipse maxima in the light curve.
Preliminary experiments indicated that the spot must be close to the orbital
plane, and so the latitude of the spot was fixed in the orbital plane for
all subsequent runs. The spot parameters that were
investigated in greater detail are (1) the longitude of the
spot centroid, (2) the spot radius, and (3) the temperature of the spot with
respect to that of the surrounding photosphere.
The results were then compared with the light curves to identify those
models that are best able to reproduce the outside eclipse light variations;
the introduction of spots were found to have only a minor impact on the
eclipse minima.

The models that best match the data favor a spot on the hemisphere of
the (cooler) secondary star facing the (hotter) primary star, but with a center that is
offset by 30 $-$ 40$^\circ$ from the line connecting the two stars - -- in
excellent agreement with the phase at which the peak Ca II emission is
seen in Figure~\ref{Kemission_phi}. In addition, large spots that cover an entire hemisphere of the
cooler star's surface and have temperatures that are
1 - 2\% hotter than the undisturbed photosphere are better
able to match the out of eclipse light variations than models with
smaller spots and hotter temperatures.

\begin{figure}[]
  \begin{center}
    \includegraphics[width=1\textwidth]{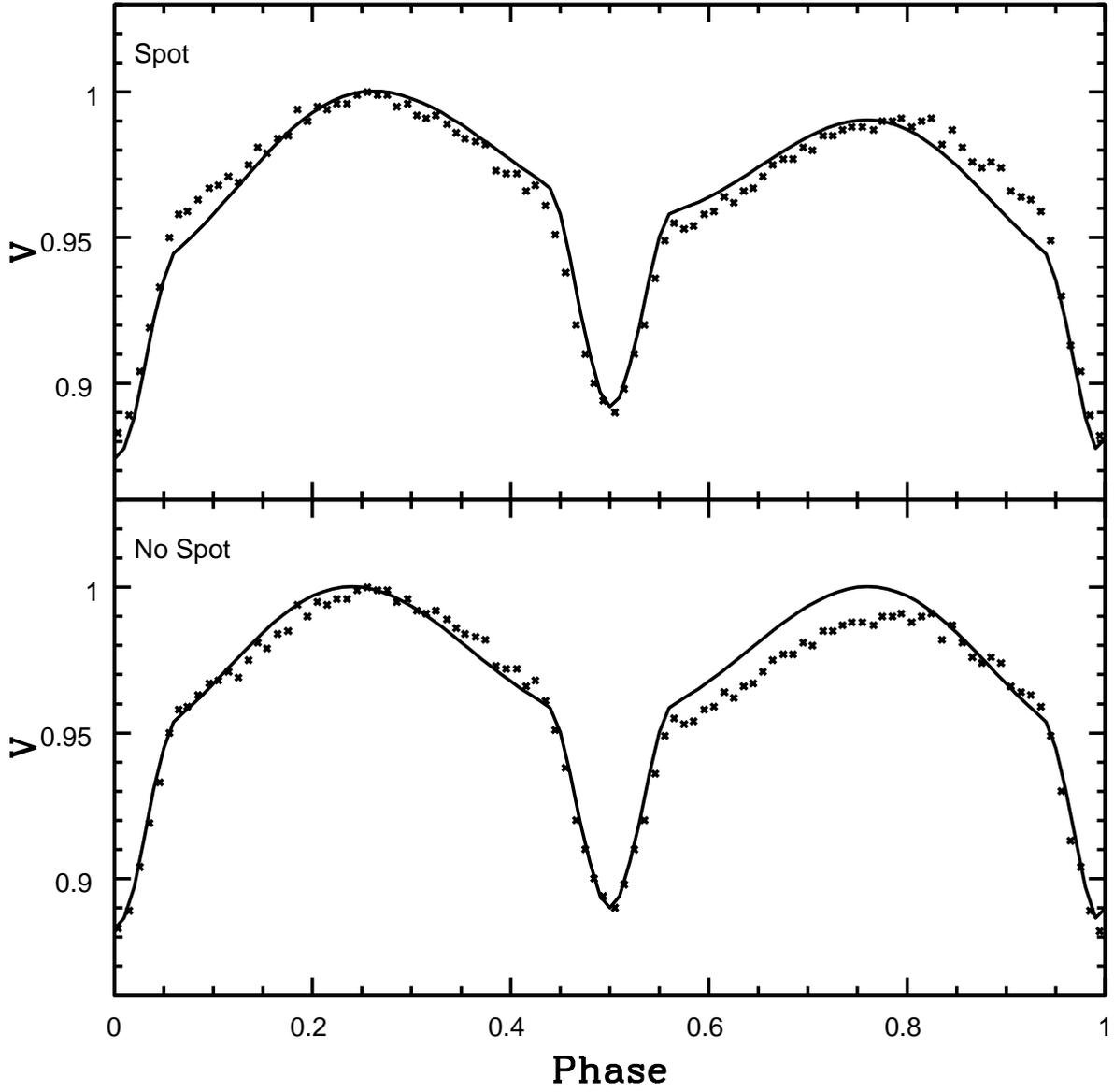}
    \caption{~~~V-band normal points compared with model light curves with (top panel)
and without (lower panel) a hot spot on the cooler star.\label{ervul_lc}}
  \end{center}
\end{figure}

The model that best matches the light curve is shown
in Figure~\ref{ervul_lc}, and the spot parameters used in this simulation are listed in Table~\ref{model_data}. The light curve generated without spots is also shown for comparison.
The agreement between the model with spots and the observations
shows systematic differences outside of eclipse, and it is conceivable
that models that include refined physics, such as multiple spots, third light
from a stream, and/or temperature structure within the spot(s) may give
better agreement with the data.

\section{Conclusions \label{summary}}

We have presented 42 high-quality Ca II H \& K spectra from the CFHT that cover nearly the entire 17-hour orbit of ER Vul in search of phase-dependent chromospheric activity. The rapid rotation causes severe line-blending and broadening such that it is difficult to distinguish between the spectra of the two stars. To do this, we invoked the Broadening Function Formalism.  We calculated BFs from the photospheric absorption lines outside of the H and K lines, convolved a sharp-lined spectrum with the BF at each orbital phase, and then subtracted the convolved template from the corresponding ER Vul spectrum.  The residuals consisted of only Ca II chromospheric emission. We focus only on the K line since the H line suffered from end effects.

The K emission appears to come from three different locations in the system:  the primary star, the secondary star and a broad stream flowing towards the secondary at a velocity of 320 km~s$^{-1}$. The in-falling material from the stream onto the secondary enhances the emission from that region indicated by a) an increase in total observed K emission and b) a hot spot on the photosphere as seen in a concurrent photometric lightcurve.  Both are consistent with an active region leading the sub-binary longitude by 30 $-$ 40$^\circ$.

ER Vul is not an ideal amplified analog of planet-induced stellar activity. Few 51 Peg primaries have spun up into sychronous rotation, thereby energizing the dynamo level seen in ER Vul and the 51 Peg secondaries are likely to be tidally locked and therefore may have less extensive magnetic fields than Jupiter. None the less, it is interesting to note that the chromospheric activity of HD179949 was synchronized with the 3.1 d revolution period of the secondary for two years (Shkolnik et al.~2003, 2004) with the active region on the primary preceding the sub-planetary point by some 0.2 in phase, similar to the leading spot on ER Vul's secondary.

%They also found that the general level of chromospheric activity on 51 Peg primaries is correlated with the projected mass of the secondary.


%% Included in this acknowledgments section are examples of the
%% AASTeX hypertext markup commands. Use \url without the optional [HREF]
%% argument when you want to print the url directly in the text. Otherwise,
%% use either \url or \anchor, with the HREF as the first argument and the
%% text to be printed in the second.

\acknowledgments

Research funding from the Canadian Natural Sciences and Engineering
Research Council (G.A.H.W., S.R., \& E.S.) and the National Research Council
of Canada (D.A.B., T.D.) is gratefully acknowledged. We are also indebted to
the CFHT staff for their care in setting up the CAFE fiber system and the
Gecko spectrograph. We would also like to thank Petr Harmanec, Russ Robb, and the paper's referee for useful discussion and commentary.

\clearpage

%% the file, uncomment the \plotone, \plottwo, and
%% If you need a layout that cannot be achieved with \plotone or
%% \plottwo, you can invoke the graphicx package directly with the
%% \includegraphics command or use \plotfiddle. For more information,
%% please see the tutorial on "Using Electronic Art with AASTeX" in the
%% documentation section at the AASTeX Web site,
%% http://www.journals.uchicago.edu/AAS/AASTeX.
%%
%% The examples below also include sample markup for submission of
%% supplemental electronic materials. As always, be sure to check
%% the instructions to authors for the journal you are submitting to
%% for specific submissions guidelines as they vary from

%% This example uses \plotone to include an EPS file scaled to
%% 80% of its natural size with \epsscale. Its caption
%% has been written to indicate that additional figure parts will be
%% available in the electronic journal.

%\begin{figure}[]
%  \begin{center}
%    \includegraphics[angle=270,width=1\textwidth]{ervul_schematic2.ps}
%    \caption[Schematic Diagram of ER Vul at $\phi$ = 0.35 and 0.50]{~~~ A schematic diagram of ER Vul at $\phi$ = 0.35 (left; as seen from above) and $\phi$ = 0.50 (right; as seen along the line of sight). The bright regions on the secondary represent the inferred hotspots while the dotted regions on the primary indicate hotspots on the unseen hemisphere of the star.\label{ervul_schematic}}
%\end{center}
%\end{figure}

\end{document}